# Pt-Decorated PdCo@Pd/C Core-Shell Nanoparticles with Enhanced Stability and Electrocatalytic Activity for Oxygen Reduction Reaction

Deli Wang,[†] Huolin L. Xin,[‡] Yingchao Yu,[†] Hongsen Wang,[†] Eric Rus,[†] David A. Muller,[‡,§] and Hector D. Abruña*,[†]

*Department of Chemistry and Chemical Biology, Cornell University, Ithaca, New York, 14853, Department of Physics, Cornell University, Ithaca, New York, 14853, School of Applied and Engineering Physics, and Kavli Institude at Cornell for nanoscale Science, Cornell University, Ithaca, New York, 14853.*

The high cost and scarcity of Pt pose serious problems for the widespread commercialization of fuel cell technologics.[1] Recently, Pd-based cathode catalysts have attracted much attention because Pd possesses similar properties (same group of the periodic table, same crystal structure, and similar atomic size) to Pt. The cost of Pd, however, is currently about one third that of Pt, and it is at least fifty times more abundant than Pt.[2] Although Pd catalysts exhibit some catalytic activity towards the oxygen reduction reaction (ORR) especially with the incorporation of other metallic elements [1c, 3] and even decorated with heteropolyacids,[4] insufficient electrocatalytic activity and stability still remain as major obstacles for fuel cell application. While the bimetallic Pt-Pd system could act as an alternative means of improving the ORR activity of Pd catalysts,[1a, 5] this kind of catalyst still requires a relatively high Pt loading.

Herein, we describe a simple method of preparing PdCo@Pd core-shell nanoparticles on carbon by post-treatment of the as-prepared PdCo/C catalysts at high temperature under flowing $H_2$. To increase the open circuit potential of the PdCo@Pd core-shell nanocatalysts for ORR, the PdCo@Pd nanoparticles are decorated with a very small amount of Pt by a spontaneous displacement reaction. Previously, nanoparticles with a Pt monolayer have been prepared by first laying down a Cu monolayer via underpotential deposition (UPD),[6] and then replacing Cu with a Pt monolayer by galvanic displacement. The amount of electrocatalyst synthesized by this method is limited because the nanoparticles must be immobilized on a glassy carbon electrode to carry out the Cu UPD. The facile method described herein is suitable for large-scale low cost production and significantly lowers the Pt loading relative to other bimetallic Pt-Pd systems.

The carbon supported PdCo@Pd core-shell nanoparticles were prepared via a two-step route (Figure S1, supporting information). First, $PdCl_2$ and $CoCl_2$ precursors, which were pre-adsorbed on the carbon, were simultaneously reduced under flowing $H_2$ in a tube furnace.[7] The as-prepared PdCo/C catalysts were then annealed at 500 °C under flowing $H_2$ to yield particles with a core-shell structure. This procedure is based on the adsorbate-induced surface segregation effect.[8] According to the literature, the surface composition of a bimetallic system can be very different from the bulk, depending on the heat of segregation and the surface mixing energy.[9] In addition, differences in the gas adsorption energy on two metals can also induce surface segregation.[10] In our case, upon annealing at high temperatures, the PdCo/C alloys undergo phase segregation, in which the Pd migrates to the surface, forming a pure Pd overlayer on the bulk alloys, since the adsorption enthalpy of H on Pd is higher than on Co.[11] The deposition of Pt was spontaneous on the PdCo@Pd surface because the equilibrium electrode potential of the $PtCl_4^{2-}/Pt$ couple (0.775 V vs SHE) is more positive than those of the $PdCl_4^{2-}/Pd$ (0.591 V vs SHE) and $Co^{2+}/Co$ (-0.28 V vs SHE) couples. Although the potential difference between the $PtCl_4^{2-}/Pt$ and the $Co^{2+}/Co$ couples is much larger than the $PtCl_4^{2-}/Pt$ and $PdCl_4^{2-}/Pd$ couples, the Pt may primarily deposit by displacing Pd since the catalyst surface is Pd-rich after anealing and prior to Pt deposition (Figure S1, supporting information). To verify that Pt spontaneously displaced Pd, we substituted the Pd/C catalyst for the core-shell PdCo@Pd/C and repeated the same procedure. The cyclic voltammetry (CV) and CO stripping behavior of the Pd/C catalyst were different after the Pt deposition procedure, suggesting that Pt was indeed deposited on the Pd/C catalysts (Figure S2, supporting information).

† Department of Chemistry and Chemical Biology, Cornell University.
‡Department of Physics, Cornell University.
‡ School of Applied and Engineering Physics, Cornell University.
§ Kavli Institute at Cornell for nanoscale Science, Cornell University.

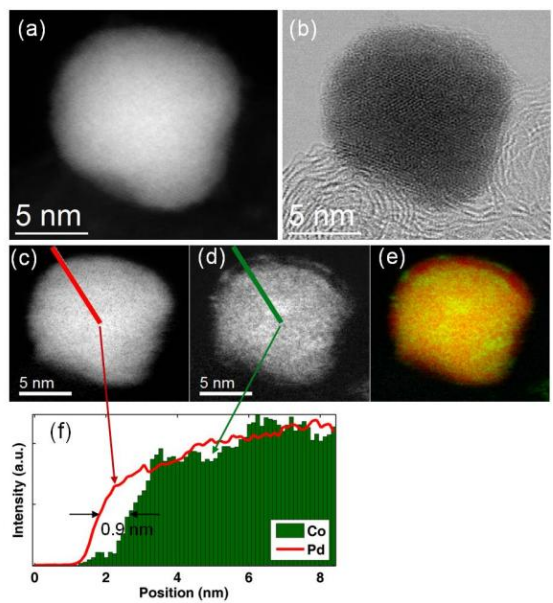

*Figure 1.* Annular dark-field (ADF) (a) and bright-field (BF) (b) STEM image of core-shell PdCo@Pd nanoparticle, EELS mapping of Pd (c), Co (d) and overlay of Pd (red) *vs* Co (green) (e), EELS line profiles of Pd and Co showing the Pd shell (f).

The core-shell structure and chemical distribution of PdCo@Pd/C nanoparticles were examined using an aberration-corrected 100 keV Nion scanning transmission electron microscopy (STEM) equipped with an Enfina electron energy loss spectrometer (EELS) where the electron beam is focused down to a 0.1-0.14 nm spot and scanned to form compositional maps.[12] Figure 1a and b show a pair of annular dark-field (ADF) and bright-field (BF) STEM images simultaneously acquired prior to EELS map acquisition showing the internal crystal structure of the nanoparticle. Figure 1c and d show the Co and Pd projected distribution within the particle extracted from a 146 x 127 pixel spectroscopic image. The Pd (red) vs Co (green) composite image (Figure 1e) demonstrates that a Pd-rich shell is formed on the surfaces. By comparing line profiles across the Pd-rich shell from the Co and Pd maps, as shown in Figure 1f, the thickness of the shell is estimated to be approximately 1 nm. A comparison of the ADF-STEM images of PdCo@Pd/C and Pt-decorated PdCo@Pd/C nanoparticles is shown in Figure S3 (supporting information). There is no appreciable distinction except a slight contrast difference between the two images.

Structural characterization of the as-prepared catalysts was carried out by X-ray diffraction (Figure S4, supporting information). All the samples showed a typical fcc pattern. After heat-treatment, the diffraction peaks of PdCo/C shifted to higher angles, relative to Pd/C, indicating a lattice contraction due to the alloying of Pd with Co. There is essentially no difference after Pt deposition, because the amount of Pt is too small to be detected using XRD. The presence of Pt, however was verified by XPS. (Figure S5, supporting information).

Figure 2 shows the polarization curves of the ORR for different catalysts (Figure 2a), and the corresponding hydrogen peroxide production measured using a rotating ring-disk electrode, in which the glassy carbon disk was modified with the catalyst and the Pt ring was used to detect the generated peroxide (Figure 2b). All electrodes were pre-treated by cycling the potential between 0.05 and 1.1 V at sweep rate of 50 mVs$^{-1}$ for 50 cycles in order to remove any surface contamination prior to ORR activity testing. The rate of the ORR is significantly enhanced at PdCo/C, and especially at the PdCo@Pd/C relative to Pd/C. The amounts of peroxide detected at the ring electrode were diminished relative to Pd/C. The CVs of PdCo@Pd/C and Pt-PdCo@Pd/C are shown in Figure 2c. The currents were normalized to the metal surface area, estimated from the coulometric charge of the oxide reduction peak. For Pt-PdCo@Pd/C, the potentials of surface oxide formation and reduction were both more positive than those of PdCo@Pd/C, suggesting the fast hydroxyl adsorption/desorption from the Pt-PdCo@Pd/C surfaces at more positive potentials. Decoration with Pt modified the ability of the catalyst for adsorbing hydroxyl species ($OH_{ad}$). The Pt decorated PdCo@Pd/C catalyst clearly had a lower OH coverage than PdCo@Pd/C over the entire potential range (Figure 2d). Since adsorbed OH species inhibit the ORR, the lower $OH_{ad}$ coverage on the surface of the Pt-PdCo@Pd/C catalysts improves the ORR kinetics. Furthermore, modification of the electronic structure could also affect the catalytic behavior of the materials. The electronic structure of Pt deposited on the PdCo@Pd/C surfaces is different from that of bulk Pt due to the so-called strain and ligand effects of the core substrate.[13]

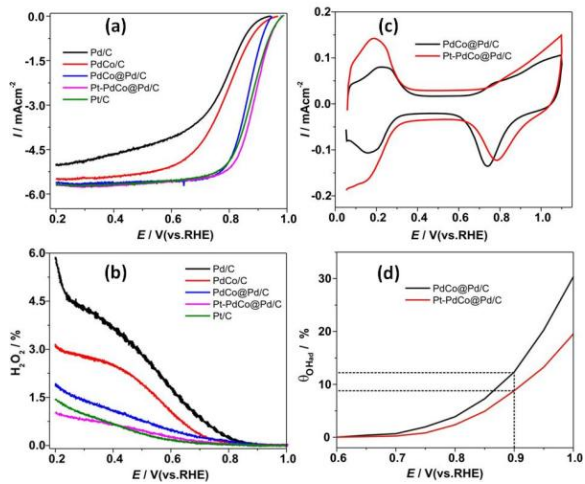

*Figure 2.* (a) ORR polarization curves. (b) peroxide generation from the ORR detected at the ring electrode in $O_2$-saturated 0.1 M $HClO_4$. sweep rate: 5 mV s$^{-1}$, rotation rate: 1600 rpm, the ring potential was held at 1.2 V (vs RHE).(c) CVs of different catalysts in 0.1 M $HClO_4$ purged with $N_2$. (d) potential dependence of hydroxyl surface. coverage($\theta_{OH}$).

Degradation of the catalysts was evaluated by cycling the electrode potential between 0.05 and 1.1 V (vs RHE). Figure 3a shows the CVs of the Pt decorated PdCo@Pd/C catalyst after an increasing number of potential cycles. For comparison, the CVs of Pd/C and PdCo@Pd/C catalysts are shown in Figure S6 (supporting information). The loss of electrochemical surface area (ECSA, obtained from the integrated coulometric charge of the oxide reduction peaks) of different catalysts after a given number of potential cycles are shown in Figure 3b. It can be seen that each catalyst shows an increased ECSA in the first few cycles, which could be attributed to surface roughening and removal of contaminants from the sample surface. However, the ECSA of Pd/C and PdCo@Pd/C catalysts decreased substantially after 50 cycles, indicating their instability. In contrast, the ECSA of the Pt decorated PdCo@Pd/C catalyst decreased only slightly after 2000 cycles, indicating that the PdCo@Pd/C catalyst was stabilized by the Pt adlayer. The Pt-PdCo@Pd/C catalyst exhibited only a 10 mV degradation in half-wave potential for the ORR after having been cycled 2000 times (Figure 3c). The methanol tolerance of the catalysts for ORR was tested in an $O_2$-saturated 0.1 M $HClO_4$ in the presence of 1 M methanol. As shown in Figure 3d, for Pt/C, the current transitioned from negative to positive at approximately +0.5 V in the presence of methanol and there was a large current peak at +0.8 V. The observed behavior was caused by the competition between methanol oxidation and oxygen reduction in which the ORR current was overwhelmed by the methanol oxidation current. However, the ORR activity of the Pt decorated PdCo@Pd/C catalyst was much less affected in the presence of methanol pointing to the potential use of this catalyst in direct methanol fuel cells.

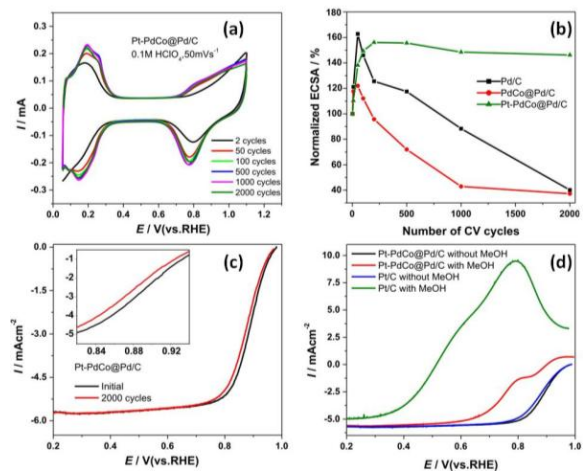

*Figure 3.* (a) CVs of the Pt decorated PdCo@Pd/C after different number of potential cycles. (b) loss of ECSA of different catalysts with number of CV cycles in $N_2$-purged 0.1 M $HClO_4$ solution. (c) ORR polarization curves of Pt decorated PdCo@Pd/C catalyst before and after cycling. (d) Polarization curves of Pt-PdCo@Pd/C and Pt/C in $O_2$-saturated 0.1 M $HClO_4$ in the presence of 1 M MeOH. rotation rate: 1600 rpm, sweep rate: 5 mV s$^{-1}$.

In conclusion, core-shell PdCo@Pd/C nanoparticles were successfully synthesized using an $H_2$-induced surface segregation effect. An ultra-low loading of Pt was deposited on the surface of the PdCo@Pd/C nanoparticles by a spontaneous displacement reaction, and the Pt-decorated PdCo@Pd/C catalyst was found to have significantly enhanced stability and ORR activity. The core-shell PdCo@Pd/C catalyst and the Pt-decprated PdCo@Pd/C catalyst showed much higher MeOH tolerance than Pt/C. The simple synthesis method and the significantly lower cost of the Pt-PdCo@Pd/C catalyst could enhance the commercial viability of fuel cell technologics.

**Acknowledgement.** This work was supported by the Department of Energy though grant DE-FG02-87ER45298, by the Energy Materials Center at Cornell, an Energy Frontier Research Center funded by the U.S. Department of Energy, Office of Science, Office of Basic Energy Sciences under Award Number DE-SC0001086. This work made use of TEM and XPS facilities of the Cornell Center for Materials Research (CCMR).

**Supporting Information Available:** Synthesis, physical characterization, and electrochemical stability test details. This material is available free of charge via the internet at http://pubs.acs.org.

**References**

(1) (a) Lim, B.; Jiang, M. J.; Camargo, P. H. C.; Cho, E. C.; Tao, J.; Lu, X. M.; Zhu, Y. M.; Xia, Y. A. *Science* **2009**, *324*, 1302. (b) Stamenkovic, V. R.; Mun, B. S.; Arenz, M.; Mayrhofer, K. J. J.; Lucas, C. A.; Wang, G. F.; Ross, P. N.; Markovic, N. M. *Nature Mater.* **2007**, *6*, 241.(c) Fernandez, J. L.; Raghuveer, V.; Manthiram, A.; Bard, A. J. *J. Am.Chem. Soc.* **2005**, *127*, 13100.
(2) Antolini, E. *Energy Environ. Sci.* **2009**, *2*, 915.
(3) (a) Savadogo, O.; Lee, K.; Oishi, K.; Mitsushima, S.; Kamiya, N.; Ota, K. I. *Electrochem. Commun.* **2004**, *6*, 105. (b) Yeh, Y. C.; Chen, H. M.; Liu, R. S.; Asakura, K.; Lo, M. Y.; Peng, Y. M.; Chan, T. S.; Lee, J. F. *Chem. Mater.* **2009**, *21*, 4030. (d) Suo, Y. G.; Zhuang, L.; Lu, J. T. *Angew. Chem. Int. Ed.* **2007**, *46*, 2862.
(4) Wang, D. L.; Lu, S. F.; Jiang, S. *Chem. Commun.* **2010**, *46*, 2058.
(5) (a) Peng, Z. M.; Yang, H. *J. Am.Chem. Soc.* **2009**, *131*, 7542.(b) Chen, Z. W.; Waje, M.; Li, W. Z.; Yan, Y. S. *Angew. Chem. Int. Ed.* **2007**, *46*, 4060.
(6) (a)Wang, J. X.; Inada, H.; Wu, L. J.; Zhu, Y. M.; Choi, Y. M.; Liu, P.; Zhou, W. P.; Adzic, R. R. *J. Am.Chem. Soc.* **2009**, *131*, 17298. (b) Zhou,W. P.; Yang, X. F. Vukmirovic, M. B.; Koel,B. E.; Jiao, J.; Peng, G. W.; Mavrikakis, M.; Adzic, R. R. *J. Am.Chem. Soc.* **2009**, *131*, 12755.
(7) (a) Wang, D. L.; Lu, S. F.; Jiang, S. P. *Electrochim.Acta* **2010**, *55*, 2964. (b) Wang, D. L.; Zhuang, L.; Lu, J. T. *J. Phys. Chem. C* **2007**, *111*,16416.
(8) Mayrhofer, K. J. J.; Juhart, V.; Hartl, K.; Hanzlik, M.; Arenz, M. *Angew. Chem. Int. Ed.* **2009**, *48*, 3529.
(9) (a) Ruban, A. V.; Skriver, H. L.; Nørskov, J. K. *Phys. Rev. B* **1999**, *59*, 15900 (b) Christensen, A.; Ruban, A. V.; Stoltze, P.; Jacobsen, K.W.; Skriver, H. L.; Nørskov, J. K.; Besenbacher, F. *Phys. Rev. B* **1997**, *56*, 5822.
(10) (a) Yin, Y.; Rioux, R. M.; Erdonmez, C. K.; Hughes, S.; Somorjai, G. A.; Alivisatos, A. P. Science, 2004, 304, 711. (b)Nerlov, J.; Chorkendorff, I. *Catal. Lett.* **1998**, *54*, 171.
(11) Popova, N. M.; Babenkova, L. V. *React. Kinet. Catal. Lett.*, **1979**, *11*, 187.
(12) Muller, D.; Fitting Kourkoutis, A. L.; Murfitt, M.; Song, J. H.; Hwang, H. Y.; Silcox, J.; Dellby, N.; Krivanek, O. L. *Science*, **2008**, *319*, 1073.
(13) (a) Karlberg, G. S. *Phy. Rev. B* **2006**, *74*.(b) Stamenkovic, V. R.; Mun, B. S.; Mayrhofer, K. J. J.; Ross, P. N.; Markovic, N. M. *J. Am.Chem. Soc.***2006**, *128*, 8813.(c) Hammer, B.; Norskov, J. K. *Adv. Catal.,* **2000**, *45*, 71.

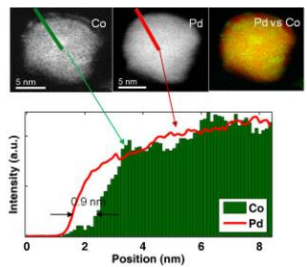

A simple method for the preparation of PdCo@Pd core-shell nanoparticles supported on carbon has been developed using an adsorbate-induced surface segregation effect. The stability and electrocatalytic activity for the oxygen reduction of PdCo@Pd nanoparticles was enhanced by a small amount of Pt, deposited via a spontaneous displacement reaction. The facile method described herein is suitable for large-scale lower cost production and significantly lowers the Pt loading and thus cost. The as-prepared PdCo@Pd and Pd-decorated PdCo@Pd nanocatalysts have higher methanol-tolerance for the ORR when compared to Pt/C, and are promising cathode catalysts for fuel cell applications.

Supporting Materials

## Pt-Decorated PdCo@Pd/C Core-Shell Nanoparticles with Enhanced Stability and Electrocatalytic Activity for Oxygen Reduction Reaction

Deli Wang, Huolin L. Xin, Yingchao Yu, Hongsen Wang, Eric Rus, David A. Muller, Hector D. Abruña*

**Experimental Section**

**Synthesis of core-shell PdCo@Pd/C nanoparticles.** The carbon supported PdCo@Pd core-shell nanoparticles were prepared via a two-step route. First, PdCo/C catalysts were synthesized using an improved impregnation method [1,2]. 30 mg of $CoCl_2$ $6H_2O$ were dissolved in 18.8 mL of 0.2 M $PdCl_2$ aqueous solution, and then a 153.6 mg of Vulcan XC-72 carbon support were dispersed in it. After ultrasonic blending for 30 min, the suspension was heated under magnetic stirring to allow the solvent to evaporate and to form a smooth, thick slurry. The slurry was dried in an oven at 60 $^o$C. After being ground in an agate mortar, the resulting dark and free-flowing powder was heated in a tube furnace at 300 $^o$C under flowing $H_2/N_2$ for 2 h. Finally, the powder was cooled to room temperature under $N_2$. The as-prepared PdCo/C catalysts were then annealed at 500 $^o$C under an $H_2$ atmosphere for 2 h to achieve the core-shell structure. Because the adsorption enthalpy of H on Pd is higher than on Co, Pd migrates to the surface forming a pure Pd overlayer on an alloy core.

**Synthesis of Pt decorated core-shell PdCo@Pd/C nanoparticles.** The Pt decorated core-shell PdCo@Pd/C nanoparticles were prepared by a spontaneous displacement reaction. 50 mg of the as-prepared carbon supported PdCo@Pd core-shell nanoparticles were suspended in 10 mL of 0.1 mM $K_2PtCl_4$ solution. The nominal atomic ratio of Pt to Pd is 1:90, which is about one monolayer of Pt deposited on the PdCo@Pd nanoparticles. After ultrasonic blending for 30 min, the suspension was then heated to 60 $^o$C under magnetic stirring and left to react for 5 h. The solid product that remained at the end of the reaction was recovered by centrifugation and dried in a vacuum oven overnight.

**Characterization.** The as-prepared catalysts were characterized by powder X-ray diffraction using a Rigaku® Ultima VI diffractometer, and diffraction patterns were collected at a scanning rate of 5 $^o$/min and with a step of 0.02 $^o$. TEM was performed using a Shottky-field-emission-gun Tecnai F20 operated at 200 kV. The electron energy loss spectroscopic maps were acquired on a 100 kV 5th-order aberration-corrected scanning transmission electron microscope (Nion UltraSTEM). When operating in high-current mode (~200 pA) for electron energy loss spectroscopic (EELS) mapping, a 1 Å point-to-point resolution is routinely achieved on this instrument (this is slightly worse than the best achievable resolution because it is operated in a source-size limited regime due to the large current used). It is difficult to quantify Co deficient areas using ADF- and BF-STEM images, because the signal is dominated by the scattering from Pd atoms. Instead, we mapped out the distribution of Pd and Co using EELS spectroscopic imaging. A 146 x 127-pixel EELS spectroscopic image of a PdCo nanoparticle was recorded over 10 minutes. The Co map was achieved by integrating over the Co L3 edge with a power-law background subtraction. In areas where the PdCo nanoparticle overlaps with the carbon support in projection, the Pd $M_{4,5}$ edge is sitting on the extended feature of the carbon K edge. Therefore, it is not reliable to use power-law background subtraction to extract the Pd M4,5 signal. Instead, the Pd map was obtained using a nonlinear least square fit of the signal by a carbon K edge and a Pd M edge reference spectrum.

**Electrochemical testing.** Electrochemical experiments were carried out in 0.1 M $HClO_4$ at room temperature using a Solartron electrochemistry station. Working electrodes were prepared by mixing the catalyst with Nafion (0.05 wt % Nafion dissolved in ethanol) solution. The mixture was sonicated and about 10.0 μL were applied onto a glassy carbon disk. After solvent evaporation, a thin layer of Nafion-catalyst-Vulcan ink remained on the GC surface to serve as the working electrode. A Pt wire was used as the counter electrode and a reversible hydrogen electrode (RHE) in the same electrolyte as the electrochemical cell was used as the reference electrode. All potentials are referred to RHE. All the electrodes were pre-treated by cycling the potential between 0.05 and 1.1 V for 50 cycles to remove any surface contamination before testing for ORR activity. The RRDE test was performed on a Pine bi-potentiostat equipped with a platinum ring and a glassy carbon disk (Pine Instrument electrode, E7R8 series). It was calibrated with and without a thin active layer on the disk by using the ferrous/ferric couple in a 10 mM $K_3[Fe(CN)_6]$ solution. The value of the collection efficiency (N) was determined to be 0.25. During the measurement, the ring potential was held at 1.2 V (vs RHE) which is sufficient to measure the current of hydrogen peroxide species diffusing from the disk, while the oxygen reduction current is negligible.

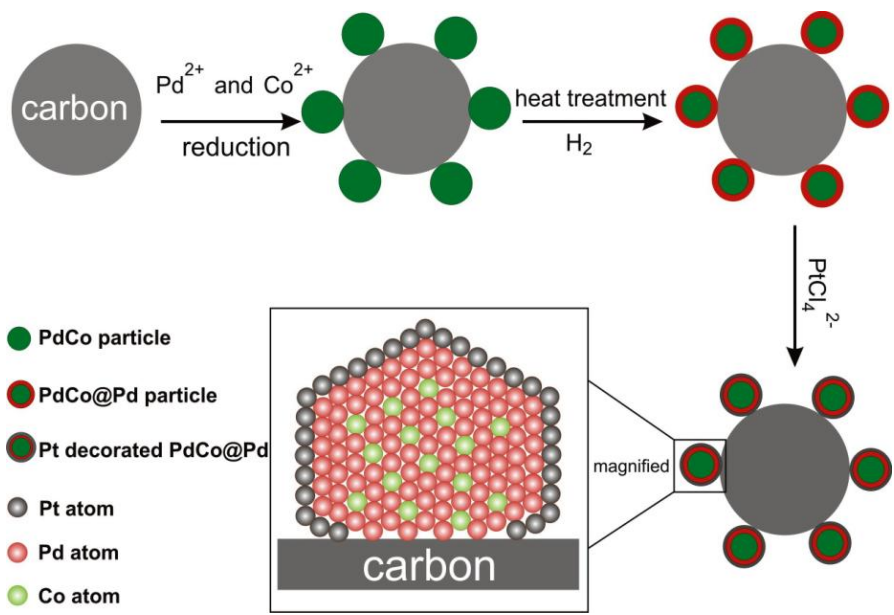

*Figure S1.* Illustration of the procedure for synthesis of core-shell PdCo@Pd/C nanoparticles and subsequent deposition of Pt

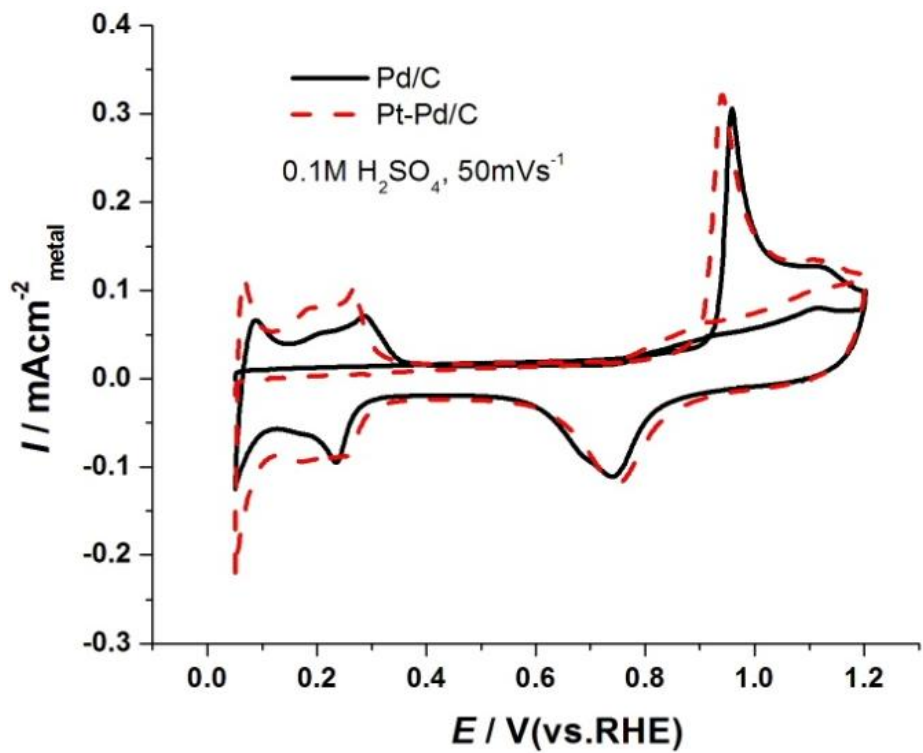

*Figure S2.* CO stripping curves of Pd/C with and without Pt deposition.

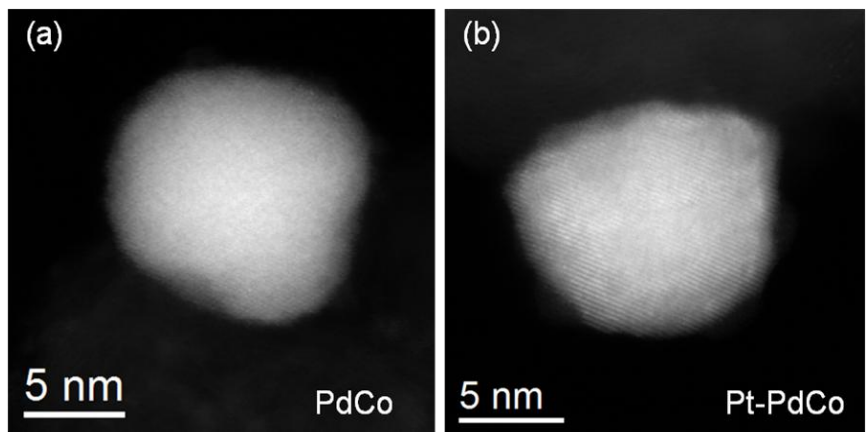

*Figure S3.* A comparison of ADF-STEM images of (a)PdCo@Pd/Cand (b)Pt-decorated of PdCo@Pd/C. There is a slight contrast difference between the two images especially at places close to the circumferences, which may indicate Pt atoms are incorporated on the surface; however, it is difficult to draw quantitative conclusions about the Pt coverage on the Pt-PdCo particle simple based upon an ADF-STEM image. This arises from the fact that the expected amount of Pt incorporated is at or below our detection limit for both ADF-STEM and EELS. Additionally, the possible existance of strain field can further complicate the interpreation of the contrast change [3].

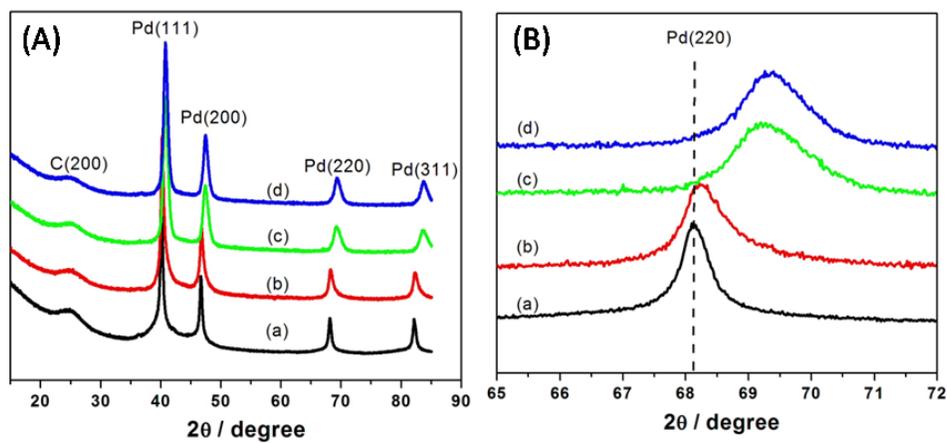

*Figure S4.* (A) XRD patterns of Pd/C(a), PdCo/C(b), PdCo@Pd/C(c) and Pt-PdCo@Pd/C(d); (B) Expanded view of the Pd(220) diffraction peaks of different samples.

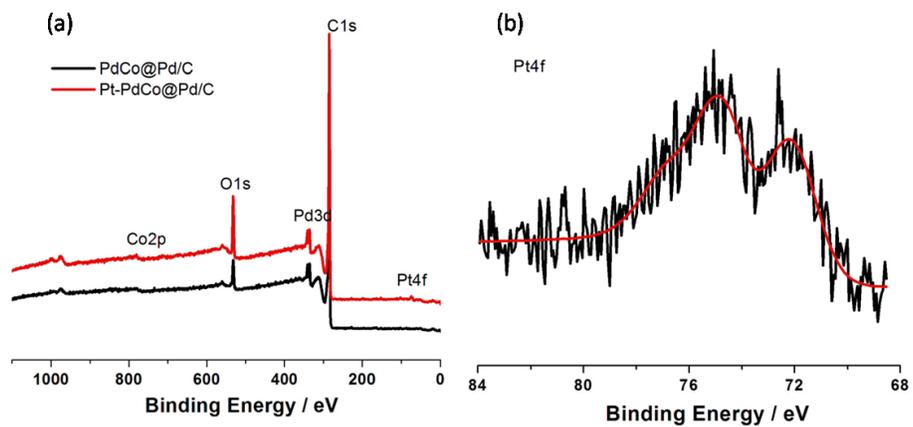

*Figure S5.* Wide-range XPS of PdCo@Pd/C and Pt-PdCo@Pd/C samples (a) and Pt 4f fine spectra (b).

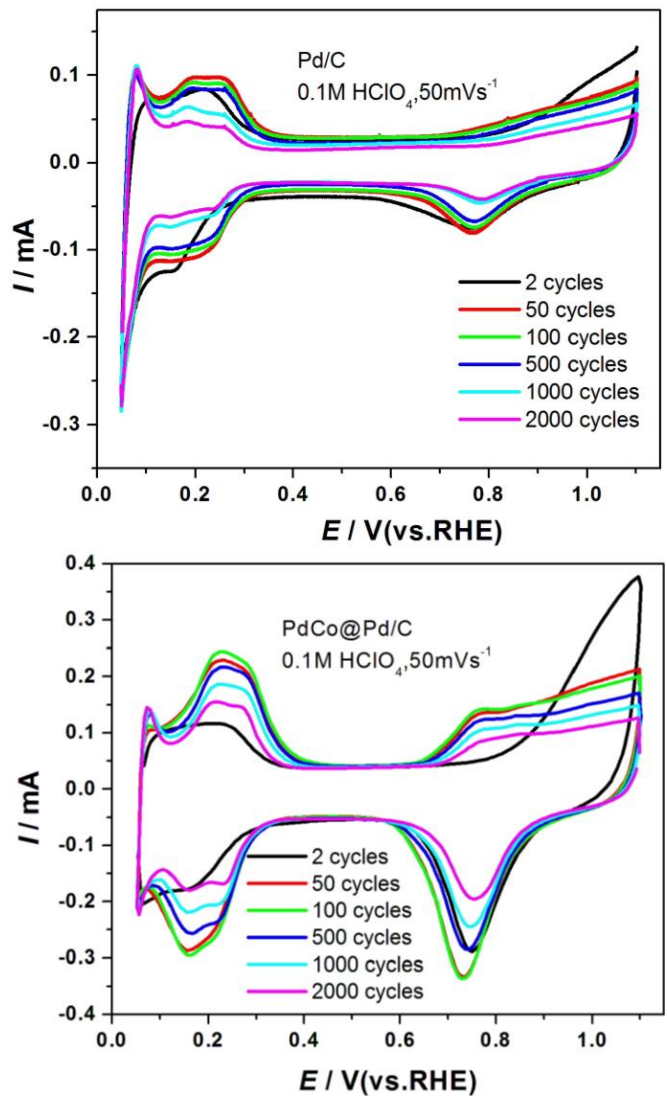

*Figure S6.* Curves of the Pd/C and PdCo@Pd/C catalyst changed with different scanning cycles.

## References

[1] D. Wang, S. Lu, S. P. Jiang, *Electrochim.Acta* **2010**, *55*, 2964.

[2] D. Wang, L. Zhuang, J. Lu, *J. Phys. Chem. C* **2007**, 111, 16416.